\documentclass[a4paper]{spie}  
\usepackage[]{graphicx}
\usepackage{amsmath,amssymb}
\usepackage{bm}

\title{Development of Uniform CdTe Pixel Detectors Based on Caltech ASIC}

\author{Kousuke Oonuki\supit{a,b}, Hokuto Inoue\supit{a,b}, Kazuhiro Nakazawa\supit{a}, Takefumi Mitani\supit{a,b}, \\Takaaki Tanaka\supit{a,b}, Tadayuki Takahashi\supit{a,b}, C. M. Hubert Chen\supit{c}, \\Walter R. Cook\supit{c}, and Fiona A. Harrison\supit{c}
\skiplinehalf
\supit{a}Institute of Space and Astronautical Science (ISAS/JAXA), Sagamihara, Kanagawa 229-8510, Japan\\
\supit{b}Department of Physics, University of Tokyo, Bunkyo, Tokyo 113-0033, Japan\\
\supit{c}Department of Physics, California Institute of Technology, Pasadena, CA 91125, USA
}

\authorinfo{Further author information: (Send correspondence to K.O.)\\K.O.: E-mail: oonuki@astro.isas.jaxa.jp, Telephone: 81 42 759 8510}

\begin{document} 
 \maketitle 

\begin{abstract}

We have developed a large CdTe pixel detector with dimensions of 23.7 $\times$ 13.0 mm${}^{2}$ and a pixel size of 448 $\times$ 448 $\mu$m$^{2}$. The detector is based on recent technologies of an uniform CdTe single crystal, a two-dimensional ASIC, and stud bump-bonding to connect pixel electrodes on the CdTe surface to the ASIC.  
Good spectra are obtained from 1051 pixels out of total 1056 pixels.  
When we operate the detector at --50 $^{\circ}$C, the energy resolution is 0.67 keV and 0.99 keV at 14 keV and 60 keV, respectively.  
Week-long stability of the detector is confirmed at operating temperatures of both --50$^{\circ}$C and --20 $^{\circ}$C.
The detector also shows high uniformity: the peak positions for all pixels agree to within 0.82 \%, and the average of the energy resolution is 1.04 keV at a temperature of --50 $^{\circ}$C.
When we normalized the peak area by the total counts detected by each pixel,
a variation of 2.1 \% is obtained.  
\end{abstract}


\section{INTRODUCTION}

One of the primary fields of high energy astro-physics
in the near future is the hard X-ray universe, where non-thermal processes 
such as particle acceleration and nucleo-systhesis become dominant.
The combination of the two new technologies of 
hard X-ray focusing mirror optics \cite{Ref:Yamashita}
and hard X-ray imaging spectrometers\cite{Ref:Takahashi_NeXT} at the focal plane
will provide two orders of magnitude improvement in both 
detection sensitivity and 
imaging resolution \cite{Ref:Kunieda,Ref:Fiona_ConX,Ref:NuSTAR}.
For example, the NeXT mission\cite{Ref:Kunieda,Ref:Proposal} proposed in Japan has mirrors with 
a focal length of 12 m and an angular resolution up to 15 arcsec\cite{Ref:Ogasaka}.
The detector is required to have an energy coverage from 5 keV to 80 keV,
with an energy resolution better than 1.0 keV (FWHM) for 60 keV line $\gamma$-rays\cite{Ref:Takahashi_SPIE2004}.
An aperture size of 20--30 mm in diameter 
with a spacial resolution of 200 -- 250 $\mu$m is needed to 
take advantage of the performance of the mirror.
The detector should have a good timing resolution 
in the range 10 -- 100 $\mu$s on an event by event basis 
to reduce the intrinsic background in the detector by the anti-coincidence technique.
Good uniformity in both the detection efficiency and the
spectroscopic properties is necessary not only to obtain high quality image
and spectra, but also to perform proper background subtraction.

Cadmium telluride (CdTe) and cadmium zinc telluride (CZT) are promising devices as the focal plane detector since they have a high detection efficiency comparable to NaI scintilators, a good energy resolution comparable to Ge detectors, and can be operated at room temperature.
Thanks to significant progress on technologies of crystal growth, large area detectors based on CdTe and CZT are now available.  
Recently, Harrison et al. has developed a large area CZT pixel detector with a newly developed low-noise ASIC for the front end\cite{Ref:Fiona_HEFT}. The detector shows a good energy resolution of $<$ 1 keV.
However, the present high pressure Bridgeman method, which is often used to grow CZT crystal, only yields polycrystals and therefore the yield of obtaining large ($>$ 1 cm$^{2}$) and uniform
portions of single CZT crystals is low. 
To control the mobility and carrier lifetime within the whole wafer seems to be difficult\cite{Ref:GSato,Ref:MSuzuki}.
This non-uniformity is the issue of the current CZT pixel detectors.
On the other hand,  CdTe crystal grown by the Traveling Heater
Method (THM-CdTe) can provide a single crystal as large as 40 mm $\times$ 40 mm \cite{Ref:Funaki}.  
Based on the THM-CdTe wafers,  we have been working on
high performance
CdTe detectors for both planar and pixel configuration\cite{Ref:Takahashi_IEEE2002,Ref:KN-NIM2003,Ref:Mitani,Ref:Tanaka}.

In order to improve the current hard X-ray imaging detector by utilizing the recent
 CdTe technology, we have developed a pixel detector under
the collaboration of ISAS and Caltech.
In this collaboration, a two-dimensional large area analog ASIC
and the read out system is prepared by Caltech.
A large size CdTe crystal with pixel electrodes and
the bump bonding technologies are prepared by ISAS.  
In this paper, uniformity of the pixel detector as well as basic performance of the CdTe pixel detector is presented.

\section{CdTe Pixel Detector} 

   \begin{figure}[b]
  \begin{center}
  \begin{minipage}[t]{7cm}
 \includegraphics[width = 6.5cm, keepaspectratio, clip]{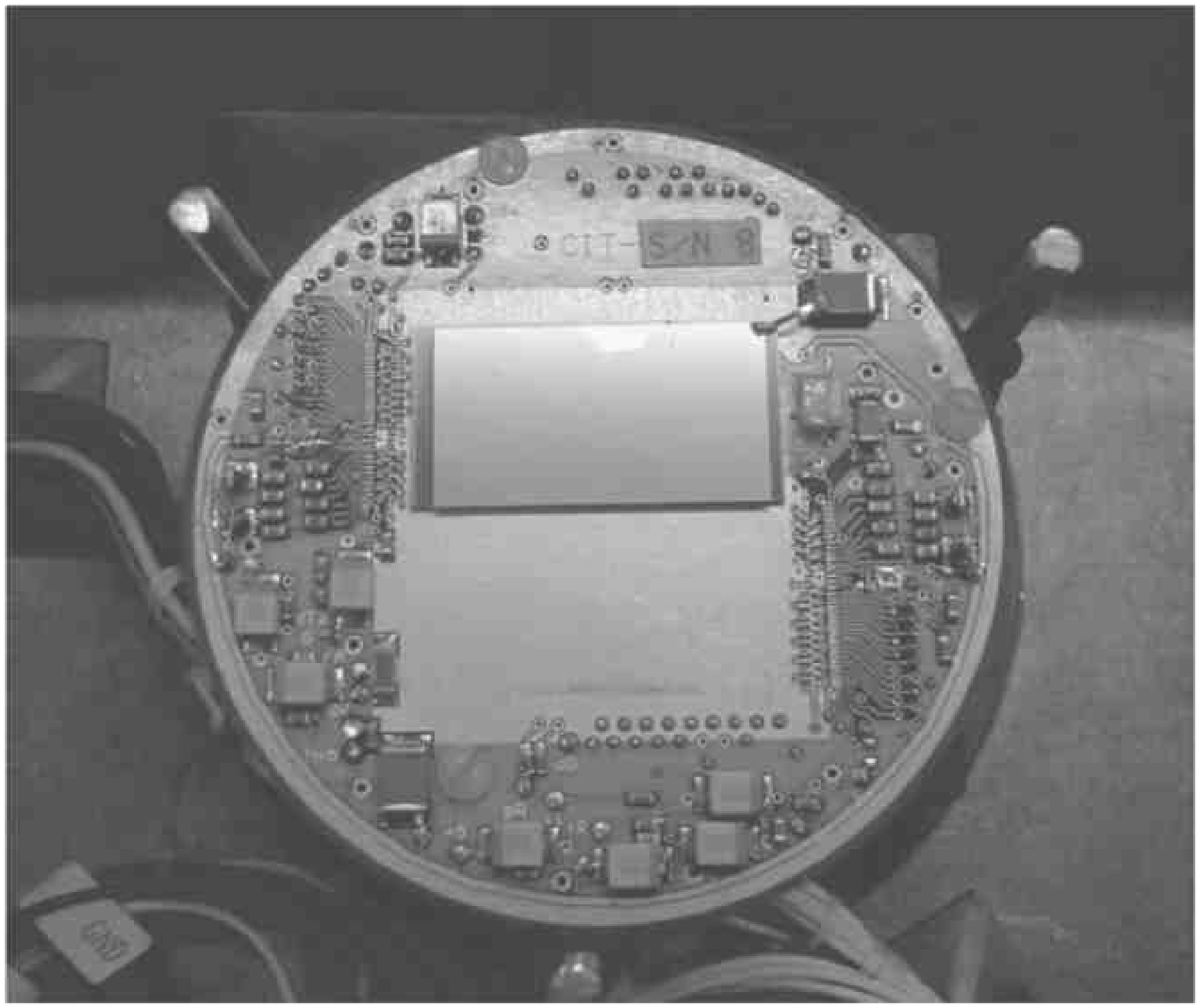}
  \caption[] {Photo of the pixel detector.  The detector has dimensions of 23.7 $\times$ 13.0 mm$^{2}$ and a thickness of 0.5 mm. The cathode surface is shown in the picture.  The bias voltage is supplied from the top board to the cathode.  }
\label{fig:detector photo}
\end{minipage}
\hspace{0.7cm}  
\begin{minipage}[t]{8cm}   
   \includegraphics[width = 7.5cm, keepaspectratio]{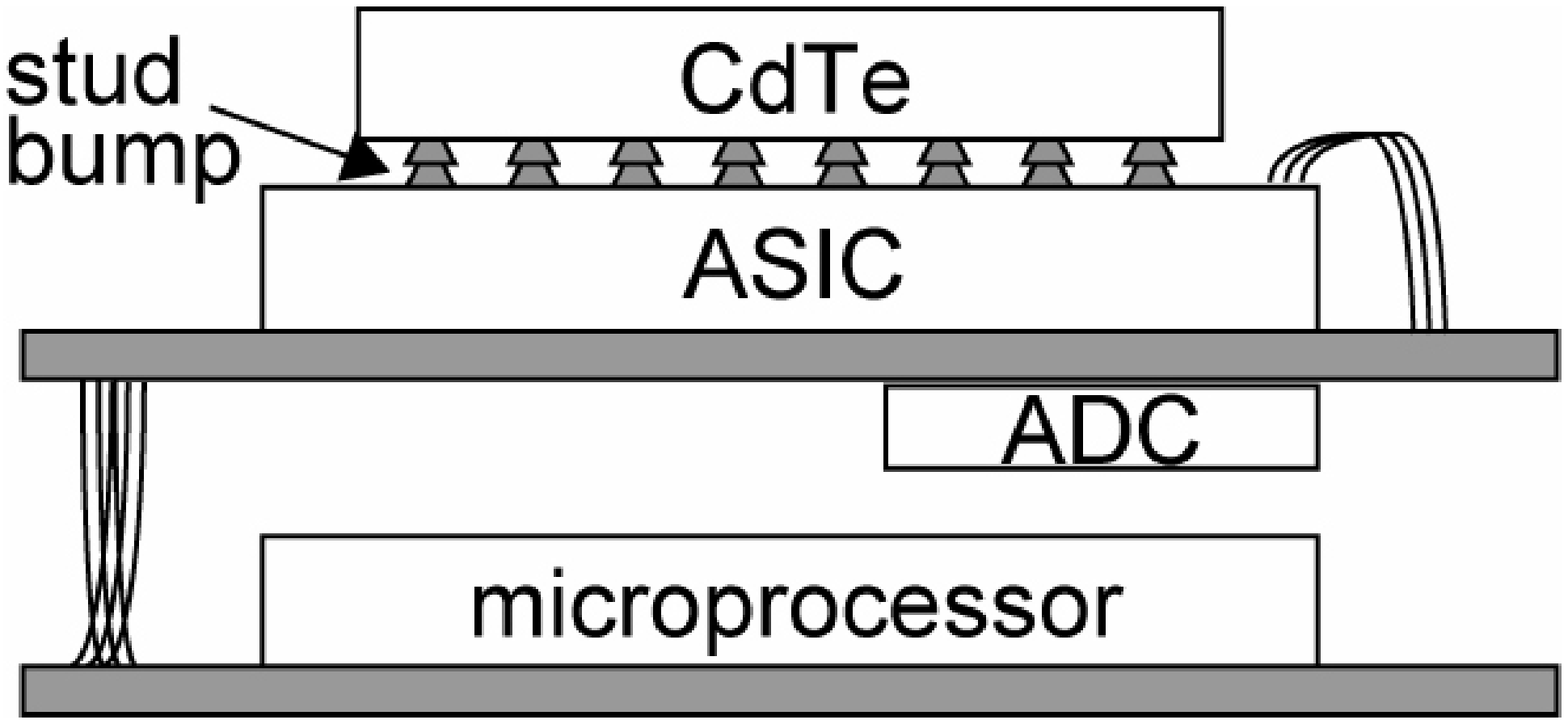}
    \caption[] {Schematic drawing.  The detector consists of CdTe, ASIC, stud bumps between them, an off-chip ADC, and a microprocessor for the readout logic.}
   \label{fig:detector drawing}
 \end{minipage}
  \end{center}
\end{figure}

The uniform charge transport properties of the wafer are very
important  for fabricating large area pixel
detectors.  The CdTe crystal used here is the THM-CdTe  manufactured by
ACRORAD\cite{Ref:Funaki}.  An electrical resistivity of $\sim$1$\times$10$^{9}$
$\Omega$ cm (p-type) is achieved by compensating the native defects
with Cl.  The grown crystal is large enough to obtain (1 1 1)$-$
oriented single crystal wafers with an area as large as 25 $\times$ 25
mm$^{2}$. We have demonstrated that the wafer shows very
good uniformity when we make a large area CdTe detector with
planar electrodes\cite{Ref:Takahashi_IEEE2002,Ref:KN-NIM2003}.  
As determined by the measurements with a large planar detector (21 $\times$ 21 mm$^{2}$ ) by using a collimated $\gamma$-ray beam
from an $^{241}$Am source, the location of the peak in the pulse height 
distribution agrees to within 0.1 \% and the
variation of the area of the 60 keV peak is less than 0.9 \%, regardless of the position.

A schematic drawing of the detector is shown in Fig. \ref{fig:detector drawing}. 
The detector consists of a CdTe wafer, a front-end ASIC and the data handling circuit including an ADC and a CPU. 
Charge signals from the individual pixel electrode are fed into the readout circuit
built in the ASIC. 
The ASIC is originally developed by the Caltech group for a CZT pixel detector
and features very low noise by means of a capacitor array \cite{Ref:Hubert_SPIE2003}.  
The power consumption of the ASIC is as low as 50 $\mu$W/pixel.

Before assembling the detector, a CdTe wafer with dimensions of 
23.7 $\times$ 13.0 mm$^{2}$ and a thickness of 0.5 mm is prepared. By using 
polished wafers with (1 1 1) orientation, Pt electrodes are formed by electroless
plating for both the common electrode (cathode) 
and pixel electrodes (anode).  The size of each pixel electrode is  448 $\mu$m $\times$ 448 $\mu$m
with  a gap of 52 $\mu$m.  No control electrode is adopted between electrodes.  A guard with 1 mm width surrounds the 24 $\times$ 44 pixels anode pixels.

 One of the most difficult parts to realize a fine pitch (finer than several
hundred microns) CdTe and/or CZT pixel detectors is to establish a
simple and robust connection technology for these fragile devices. 
High compression and/or high ambient temperature
would damage the crystal.  Also the co-planarity of the CdTe
wafer is measured to be 2 $\mu$m at most, which is much worse than
that of usual silicon wafers.  Usual indium-ball soldering might not
be appropriate for this purpose. Also indium is easily oxidizable, and
needs flux which could contaminate the detector surface. 
Recently, we have developed indium and gold (In/Au) stud-bump bonding technology, optimized for CdTe detectors, with Mitsubishi Heavy Industries (MHI) in Japan. By
following the prescription given in Takahashi et al. \cite{Ref:Takahashi_IEEE2001}, but
with a slightly modified bump condition, we connected the CdTe wafer to the ASIC.
In order to
minimize the effect of incomplete charge collection, we apply a negative bias on the common electrode. 
After we assembled the detector, we 
have confirmed the connection of 1055 pixels out of 1056 pixels. There is one bad pixel
which shows very high noise, regardless of any conditions, such as bias voltage, and temperature.  Currently, we do not understand whether the read out circuit is broken, or the bump is not connected for this particular pixel.

\begin{figure}[]
   \begin{center}
  \includegraphics[width = 17cm, keepaspectratio, clip ]{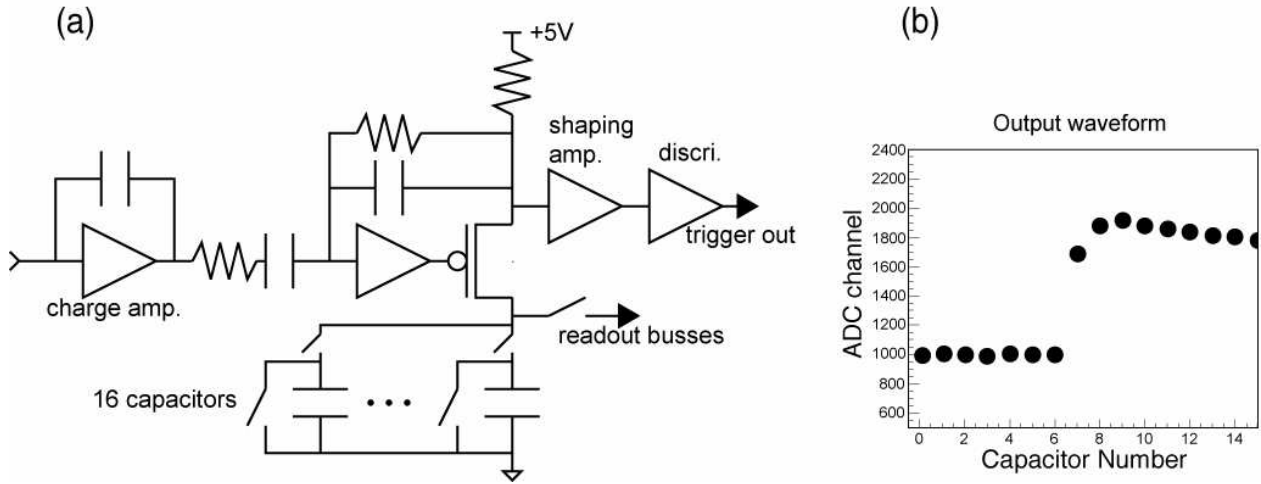}
   \end{center}
   \caption[] {(a) Schematic diagram of the the circuit for one pixel which is implemented in the ASIC,
 and (b) An example of an output waveform.  The circuit has a charge amplifier, a shaping amplifier, a discriminator, and a capacitor array of 16 capacitors.  The outputs are sixteen samplings by the array of capacitors.}
\label{fig:ASIC architecture}
 \end{figure}

In the ASIC (Fig. \ref{fig:ASIC architecture} (a)), output signal from the preamplifier is continuously integrated by one capacitor  to the next with a time interval of 1 $\mu$s. After a trigger signal is issued from a comparator, the switching sequence continues for the next eight capacitors and then the sequence is freezed for the subsequent readout.  
The charge in sixteen capacitors is sequentially read out  as voltage and converted by the
external ADC. Figure \ref{fig:ASIC architecture} (b) shows an example of the output signal for one pixel. Pulse height information for each X-ray photon is calculated by analyzing the recorded  pulse shape by the filter algorithm implemented in DSP or by software.
The ASIC also reads out data from the surrounding eight pixels together with the data from the triggered pixel. A detailed description of the ASIC is given in Chen et al\cite{Ref:Hubert_SPIE2003}. 
For the detector used in this experiment, triggers from five pixels are turned
off, because these channels show rather high noise as compared to other channels. The typical noise level is about 50 ADC channel for normal pixels, while the noise level of bad channels is 
larger than 200 ADC channel.  From hereafter, we analyze spectra
 from 1051 pixels out of total 1056 pixels.  At $-$20  $^{\circ}$C, the FWHM of the peak measured 
by the test pulse 
ranges from  0.4 keV to 0.6 keV when the bias voltage is not applied. This corresponds to the electronic noise of 40 -- 60 $e^{-}$
(RMS).
 

The spectrum from each pixel is constructed from the pulse height information
associated with the trigger. There are two types of events; 1) single-pixel events
in which the pulse height of pixels surrounding the triggered pixel is comparable
with the noise level, and 2) multi-pixel events in which the energy deposited in the CdTe
detector is shared with multiple pixels.

\section{Signal Processing and Spectral Analysis}

In order to study the spectral performance of the CdTe pixel detector, we start from the search for the optimum operating condition and the 
investigation of the response from a single pixel. 
The overall performance, including the pixel by pixel variation
will be discussed in the next section.

\begin{figure}[b]
  	\begin{center}
		  \includegraphics[width = 7cm, keepaspectratio, clip]{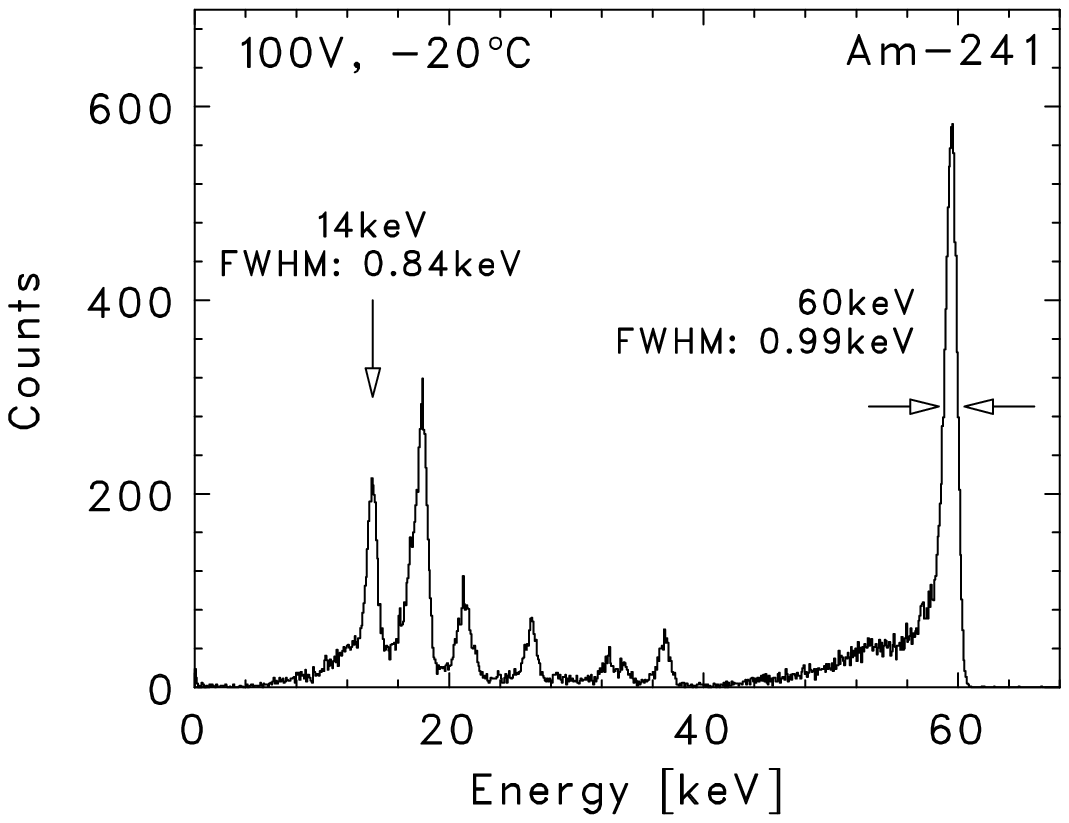}
		  \hspace{0.7cm}
  		\includegraphics[width = 7cm, keepaspectratio, clip]{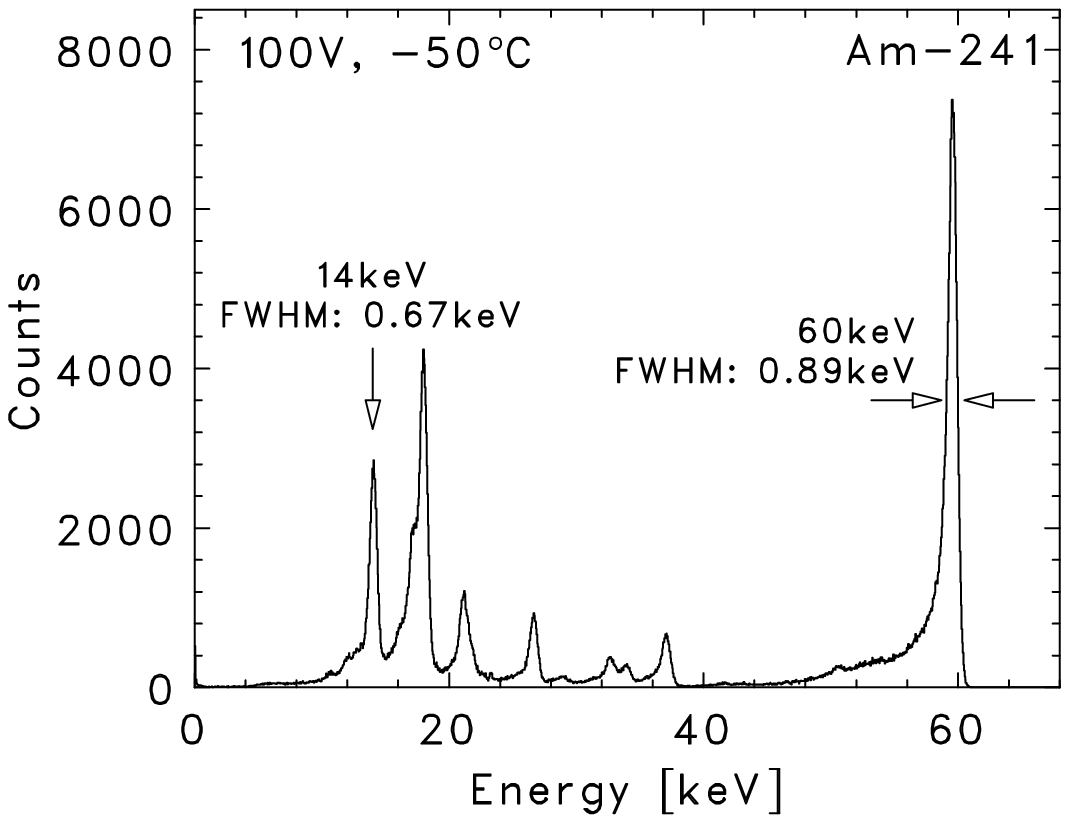}
 \end{center}
		\caption[] {$^{241}$Am spectra obtained at --20 $^{\circ}$C and --50 $^{\circ}$C.  $^{241}$Am illuminates the detector  from 40 cm above the cathode plane of the detector.   }
		\label{fig:Am spectra}
\end{figure}

As the first demonstration of the operation of the CdTe pixel detector, Fig. \ref{fig:Am spectra} shows $^{241}$Am
spectra taken at $-$20  $^{\circ}$C and $-50$  $^{\circ}$C. 
Single-pixel events are used to make the spectra.
Energy resolutions better than 1 keV (FWHM) are obtained for $\gamma$-ray lines from 14 keV to 60 keV in the spectra. 
The best resolution is obtained when the detector is operated at the lowest temperature, $-$50  $^{\circ}$C. 
With a bias voltage of 100V, the energy resolution
at 14 keV and 60 keV is 0.67 keV and 0.89 keV, respectively.
 The improvement in the spectral resolution is due to the reduction of the leakage current.   
At $-$20  $^{\circ}$C, the leakage current is estimated to be
 an order of 10 pA per pixel at --100 V, which is calculated from the results obtained with a 
 planar detector with Pt/CdTe/Pt electrode configuration. 
 Although we need to perform more quantitative measurements, the leakage current at --50 $^{\circ}$C
is expected to be reduced by almost an order of magnitude as compared to that at --20 $^{\circ}$C.

\begin{figure}[p]
\begin{center}
	\begin{minipage}[t]{14cm}
	\begin{center}
		  \includegraphics[height=9cm, keepaspectratio, clip]{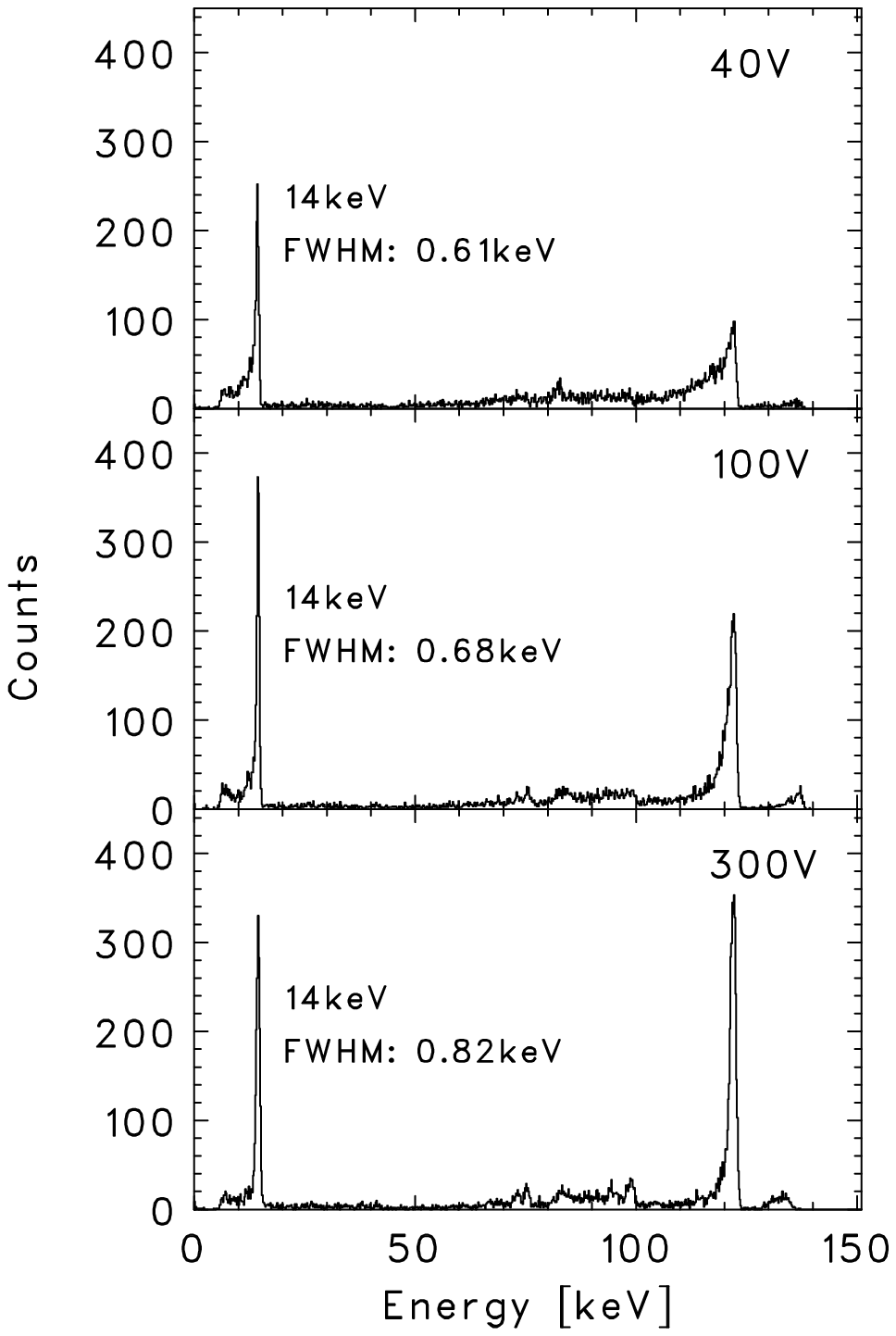}
		\caption[] {$^{57}$Co spectrum obtained at an operating temperature of --50 $^{\circ}$C with different bias voltages.  The applied bias is 40 V, 100 V, and 300 V.  All the spectra are drawn by using single-pixel events in which no signal larger than 5 keV is detected in the surrounding pixels.  The energy resolution for the 14 keV line is as low as 0.61 keV at the bias voltage of 40V.  The size of the low-energy tail is reduced as the bias voltage increases.} 
		\label{fig:spectra Co57 with diff bias} 
	\end{center}
	\end{minipage}

\vspace{0.7cm}
%
	\begin{minipage}[t]{8cm}
	\begin{center}
	\includegraphics[height=6.4cm, keepaspectratio, clip]{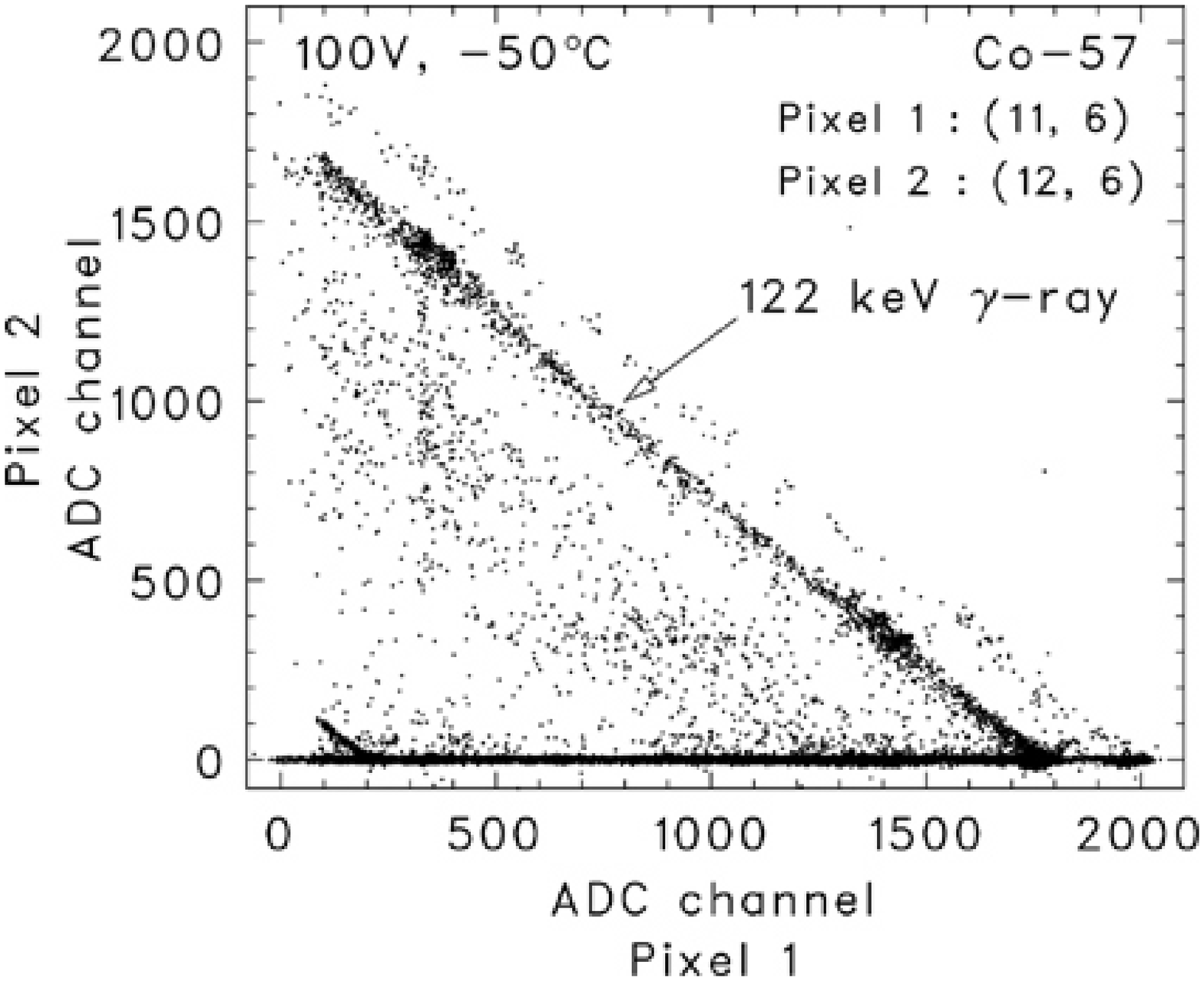}
		\caption{Correlation between pulse height (ADC channel) from a triggering pixel and that of the adjacent pixel.  The horizontal axis is the ADC channel for the triggered pixel while the vertical axis for the adjacent pixel.}  
		\label{fig:hist2dim split}
	\end{center}
	\end{minipage}
\hspace{0.7cm}
	\begin{minipage}[t]{8cm}
	\begin{center}
		  \includegraphics[width=7.5cm, keepaspectratio, clip]{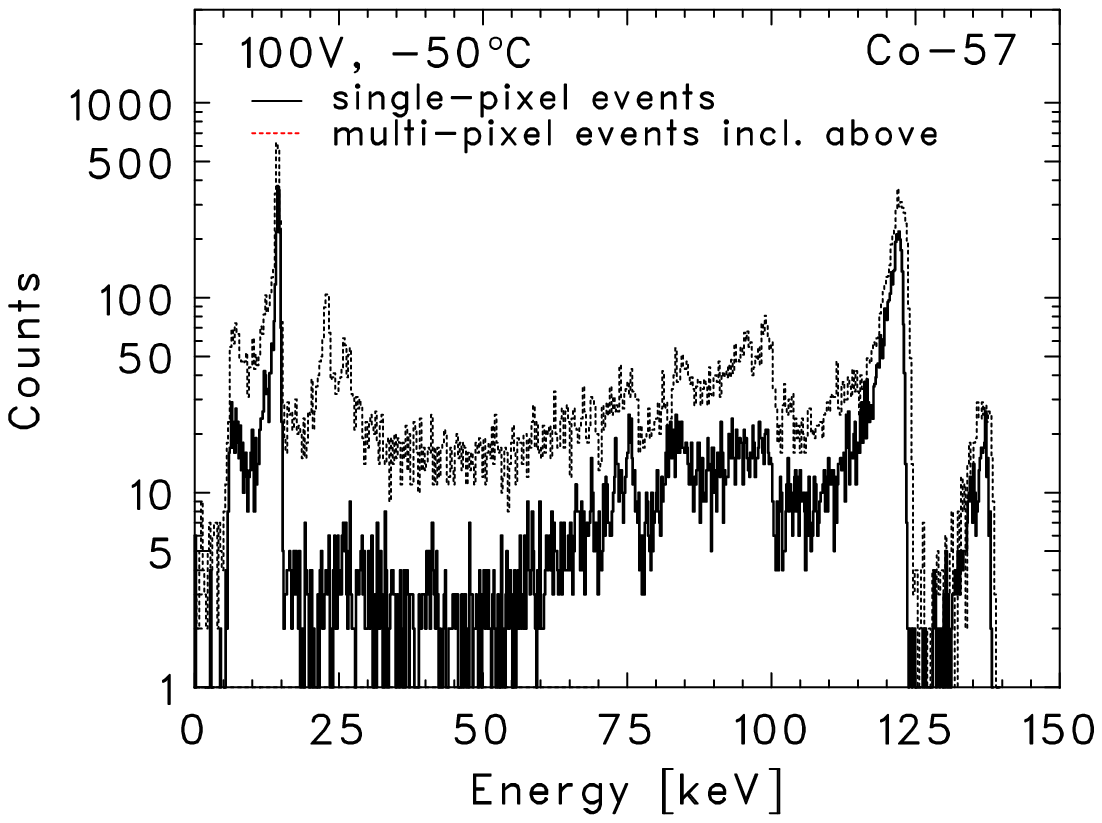}
		\caption{Spectra both for single-pixel events (black) and multi-pixel events including single-pixel events (dash).  A low energy threshold of 5 keV is used to discriminate hits in the surrounding pixels.}
		\label{fig:spectra single & split}
	\end{center}
	\end{minipage}
%
\end{center}
\end{figure}

Charge collection efficiency (CCE) is an important issue for CdTe and CZT, because
electrons and holes generated in these semiconductor have
 low mobility and a short life-time\cite{Ref:Takahashi_IEEE2001_2}. 
The low CCE results in the tail-like structure below the peak in the spectrum,
which is actually seen in the $^{241}$Am spectra (Fig. \ref{fig:Am spectra}).
We study the CCE quantitatively by measuring the $^{57}$Co spectrum at
 different bias voltages under an operating temperature of $-$50  $^{\circ}$C.
As clearly shown in the case of the 40 V operation (Fig. \ref{fig:spectra Co57 with diff bias}), 
a significant amount of charge loss is seen as a low-energy tail of the 122 keV line. 
As the bias voltage is increased to 300 V,
a sharp 122 keV peak is obtained.  
A higher bias voltage is better in terms of the CCE because the fraction of electrons and holes that reach the electrodes increases.
However, high bias voltage deteriorates the spectral resolution due to the increase of the leakage current.
In the measurements, the best energy resolution for the 14 keV line is obtained at 40V. 
On the other hand, the best energy resolution of 1.47 keV for the 122 keV line is obtained at 300 V.
This is because most of the 14 keV photons interact close to the cathode electrodes,
and because the transit of electrons, which is less affected by charge loss, becomes responsible for 
the pulse height.  As an optimum value, we use 300 V as the standard bias voltage from here after.

Since the pixel size is as small as 500 $\mu$m for the CdTe pixel detector,
 we need to investigate the charge splitting among the adjacent pixels
(multi-pixel event).
Figure \ref{fig:hist2dim split} shows a correlation  of pulse heights between a triggering pixel and one of the adjacent pixels. It is shown that
the energy deposition, corresponds to the 122 keV
$\gamma$-ray, is shared between two pixels, for some fraction. 
%
%
Figure \ref{fig:spectra single & split} compares spectra for  single-pixel and multi-pixel
events.  It is clearly shown that the amount of the tail component is more pronounced when
we include multi-pixel events. This implies that the selection of single-pixel events is important
to obtain ``tail-less" spectra in the CdTe pixel detector.

\begin{figure}[thbp]
\begin{center}
\includegraphics[width=6.4cm, keepaspectratio, clip]{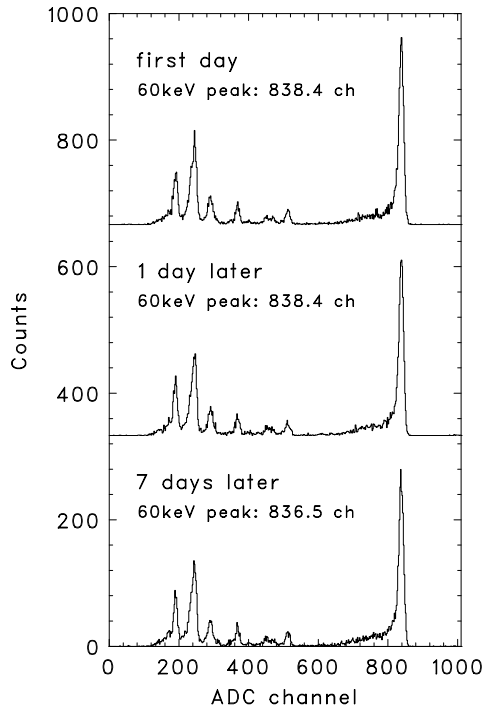}
	\caption{Spectra taken at the first day, 1 day later, and a week later after the bias voltage is applied.  An operating temperature is --50 $^{\circ}$C and a bias voltage is 300 V.  The spectra are drawn by summing 10 pixels.}
	\label{fig:stability}
\end{center}
\end{figure}

Long-term stability is an important issue for CdTe and CZT semiconductors
\cite{Ref:Takahashi_IEEE2001}.  As discussed in our previous publication\cite{Ref:Takahashi_IEEE2002}, 
a CdTe detector with Pt electrode for both cathode and
anode sides (Pt/CdTe/Pt) shows good stability. We do not see any spectral degradation
for a long term operation of a week or more. In order to verify this with the
CdTe pixel detector, we operate it
for seven days  at an operating temperature of --50 $^{\circ}$C under a bias of 300 V.
An $^{241}$Am source is placed above the detector during the measurement. 
Figure \ref{fig:stability} shows the spectra taken at the first day, 1 day later, and 7 days later.
No degradation can be observed in the peak positions and the energy resolution.  
The location of the peak in the pulse height distribution agrees to within 0.2\% . 
The energy resolutions (FWHM) agree within statistical errors.
The detector also shows a stable spectral performance at --20 $^{\circ}$C over one week.

\section{Uniformity of the Detector}

\begin{figure}[p]
\begin{center}
	\begin{minipage}[t]{14cm}
	\begin{center}
		\includegraphics[width=12cm, keepaspectratio, clip]{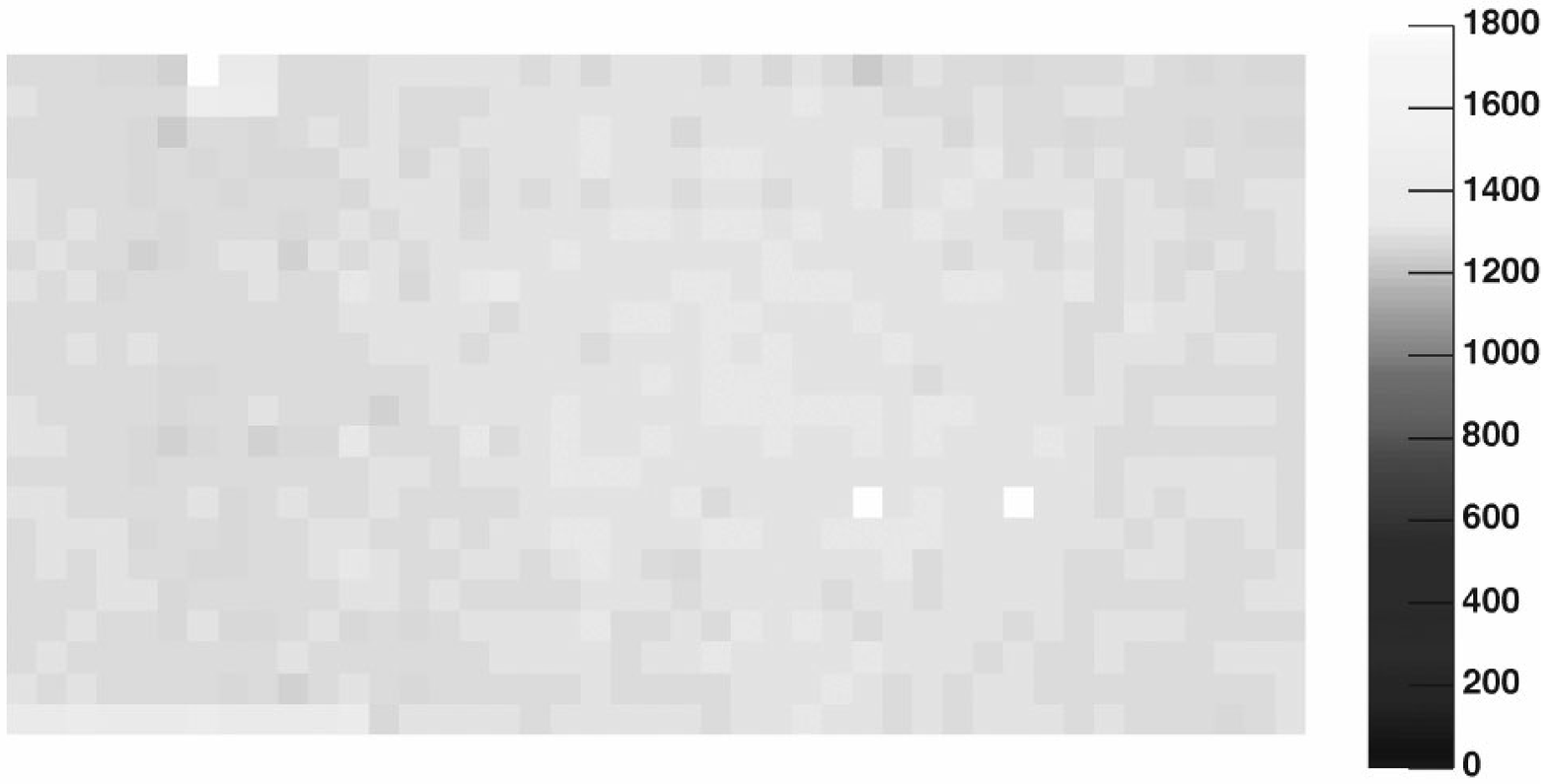}
		\caption{Map of the peak positions of the 60 keV line.  The white spots at the right center and top left are disabled pixels.}
		\label{fig:peak pos 2dim} 
	\end{center}
	\end{minipage}
\end{center}
\begin{center}
	\begin{minipage}[t]{14cm}
	\begin{center}
		\includegraphics[width=9cm, keepaspectratio, clip]{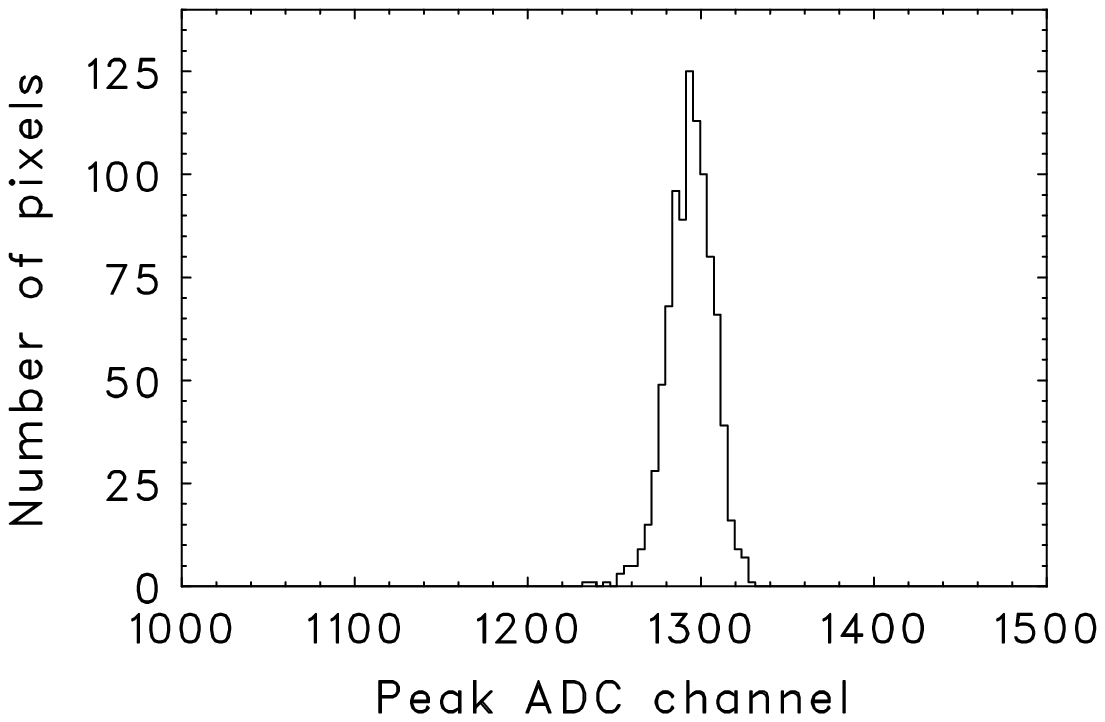}\hspace{1cm}
		\caption{Distribution of the 60 keV peak positions after the analog gain of the ASIC is calibrated by means of pulser data.  The variation for the peak position is less than 0.82 \%.}
		\label{fig:peak pos 1dim}
	\end{center}
	\end{minipage}

\vspace{0.5cm}
	\begin{minipage}[t]{14cm}
	\begin{center}
		\includegraphics[keepaspectratio, width = 9cm, clip]{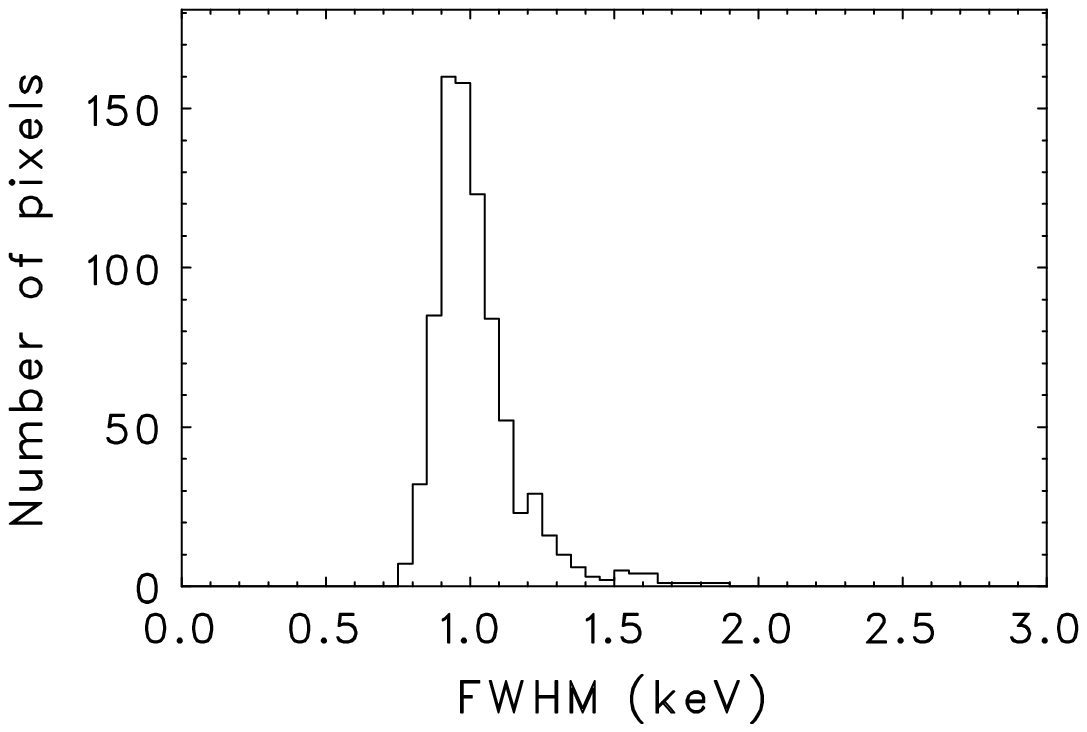}\hspace{1cm}
		\caption[] {Distribution of the energy resolution (FWHM) for the 60 keV line.  The mean and standard deviation of the distribution are 1.04 keV, and 0.10 keV.}
   		\label{fig:de} 
	\end{center}
	\end{minipage}
\end{center}
\end{figure}

Evaluating the detector uniformity response is an important objective of this experiment.
We analyze the uniformity of the detector in terms of spectral properties such as the peak position
and the spectral resolution, as well as the detection efficiency. 

The location of the peak in the spectrum for mono-energetic $\gamma$-rays has a strong
correlation with the charge transport properties of the material\cite{Ref:Takahashi_IEEE2001_2,Ref:GSato}, especially
for electrons. If the mobility-lifetime product is not uniform in the wafer used in the detector, the
peak position changes with the pixel location, since we apply a 
single bias voltage over the entire
detector plane. The variations of the size of low-energy tail is explained if the transport 
properties for holes have some spatial distribution. These spatial variations of charge transport
properties could be introduced if the crystal is not uniformly grown.

We study the uniformity by irradiating with an $^{241}$Am source from 40 cm above the detector, and obtain flat illumination data. 
The detector is operated under a bias of 300 V, at a temperature of $-50$ $^{\circ}$C.
In order to evaluate  the charge transport properties, the analog gains for each pixel are calibrated by the using the on-chip test pulse. 
The average gain variation is 2.8 \%, 
 consistent with the number reported in Harrison et al\cite{Ref:Fiona_SPIE2003}.
Figure \ref{fig:peak pos 2dim} shows the spatial distribution of the peak position in ADC channel after
the gain of readout circuit in the ASIC is calibrated.
Distribution of the peak position is almost flat and shows a standard deviation of 0.82 \%
(Fig. \ref{fig:peak pos 1dim}), which agrees with the result from the scanning experiment of a planar CdTe detector \cite{Ref:Takahashi_IEEE2001,Ref:KN-NIM2003}. 
Distribution of the energy resolution is also presented in Fig. \ref{fig:de}. 
The average energy resolution is 1.04 keV, and the standard deviation is 0.10 keV.

\begin{figure}[b]
\begin{center}
	\begin{minipage}[t]{7cm}
	\begin{center}
		\includegraphics[keepaspectratio, width = 7cm, clip]{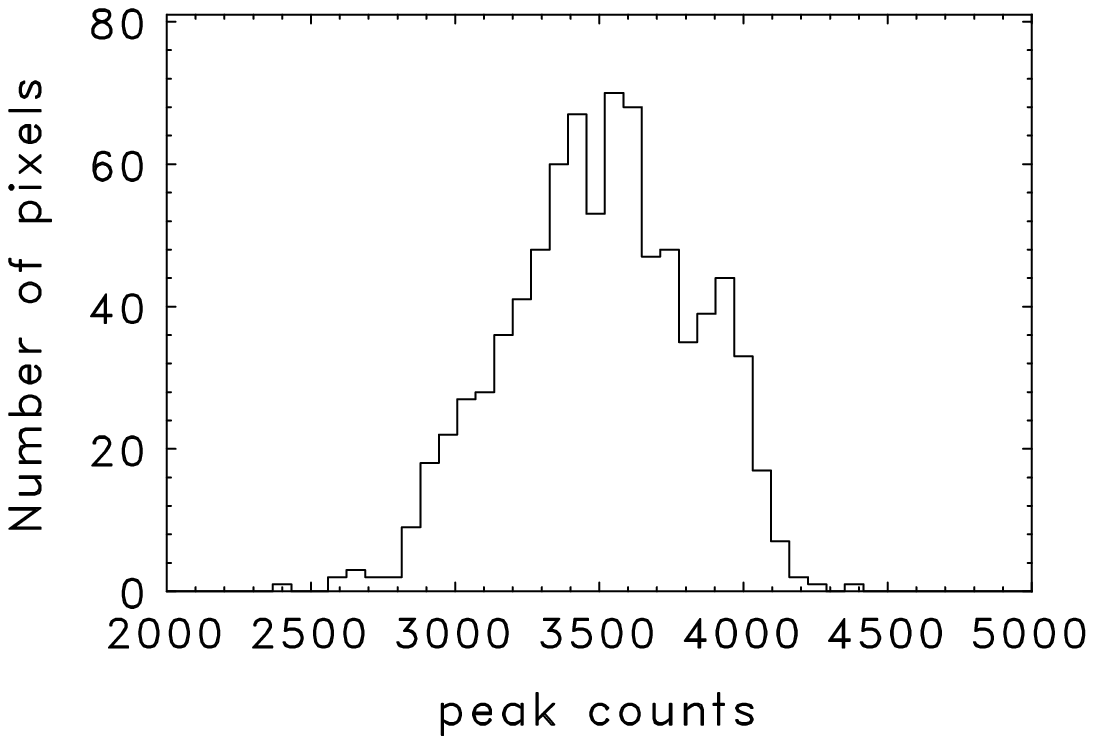}
		\caption{Count variation for the 60 keV peak.  The variation is 8.9 \%.}
		\label{fig:60keV counts}
   	\end{center}
	\end{minipage}
\hspace{0.7cm}
	\begin{minipage}[t]{7cm}
	\begin{center}
		\includegraphics[keepaspectratio, width = 7cm, clip]{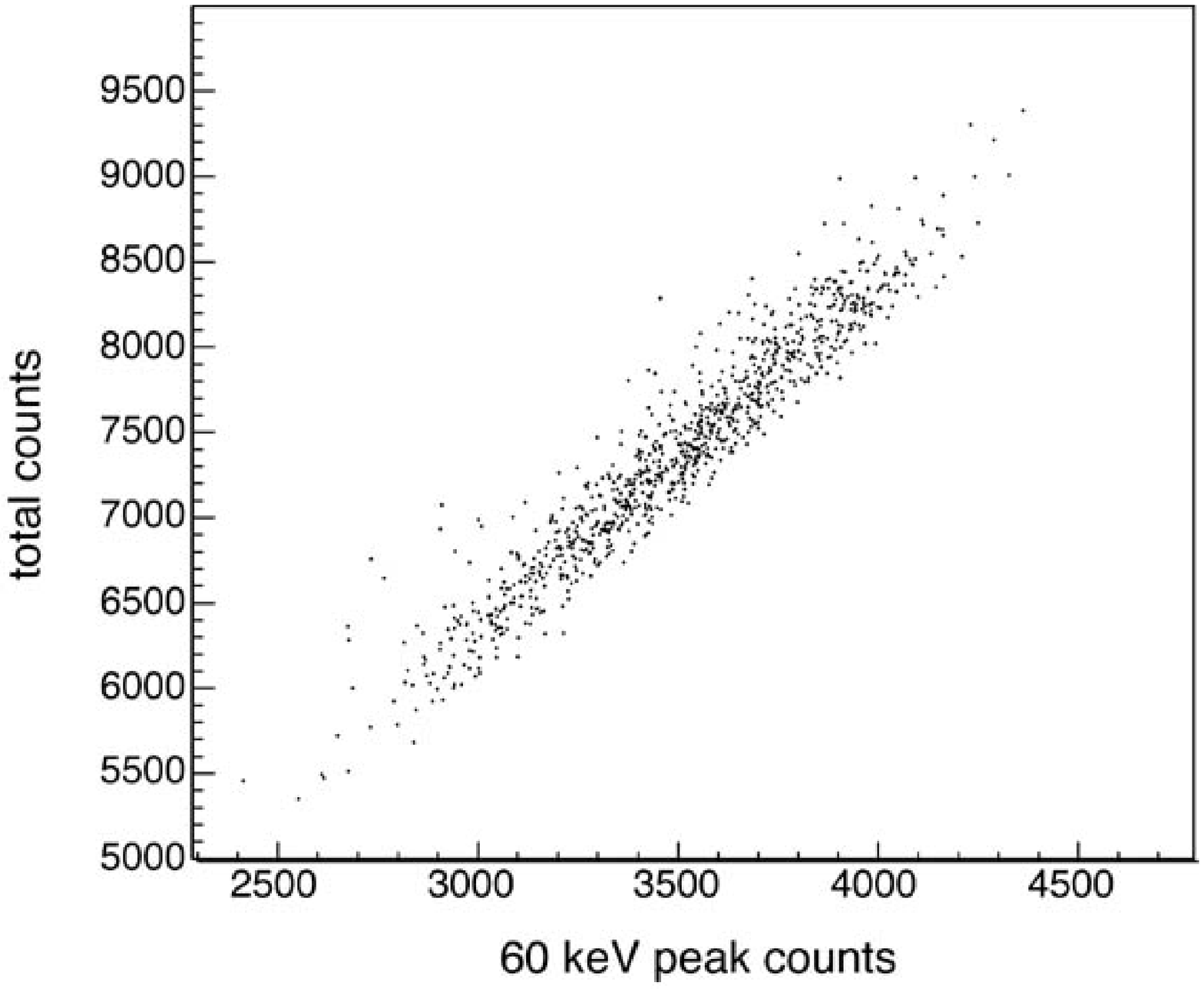}
		\caption{Correlation between 60 keV peak counts and total counts for each pixel.}
    	\label{fig:60keV counts vs total counts} 
	\end{center}
	\end{minipage}
	
\vspace{0.7cm}
	\includegraphics[keepaspectratio, width = 6.5cm, clip]{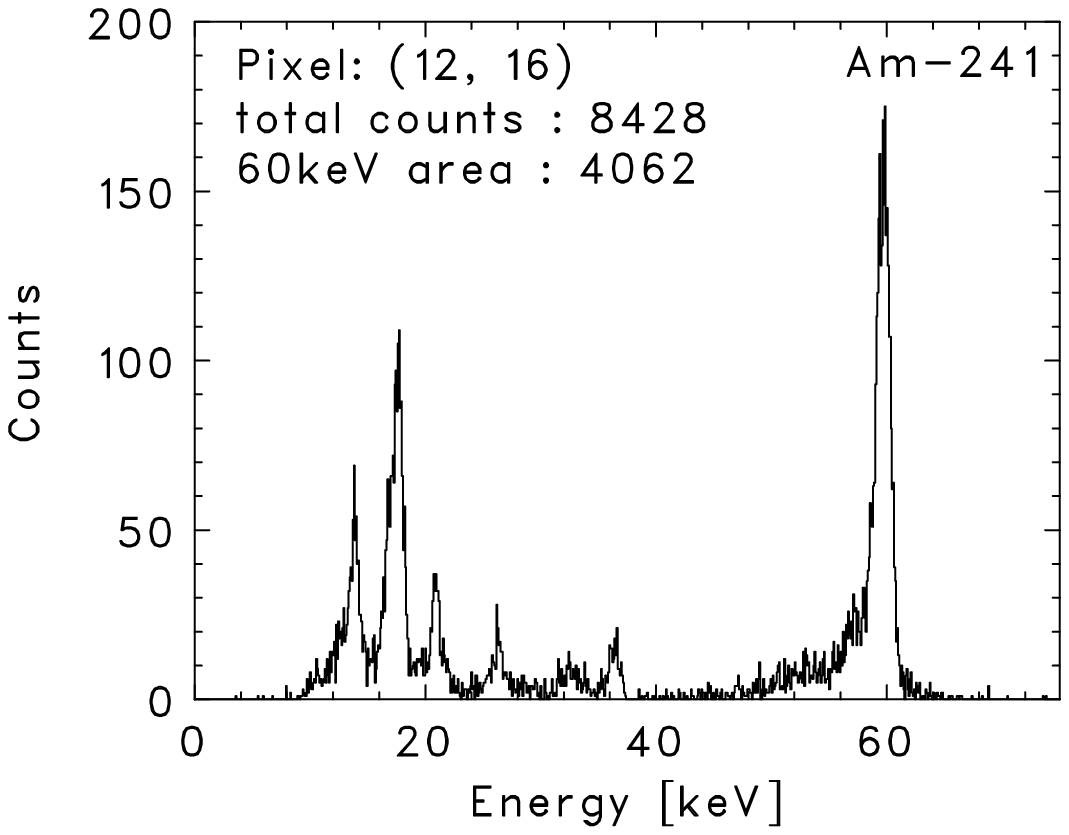}
	\includegraphics[keepaspectratio, width = 6.5cm, clip]{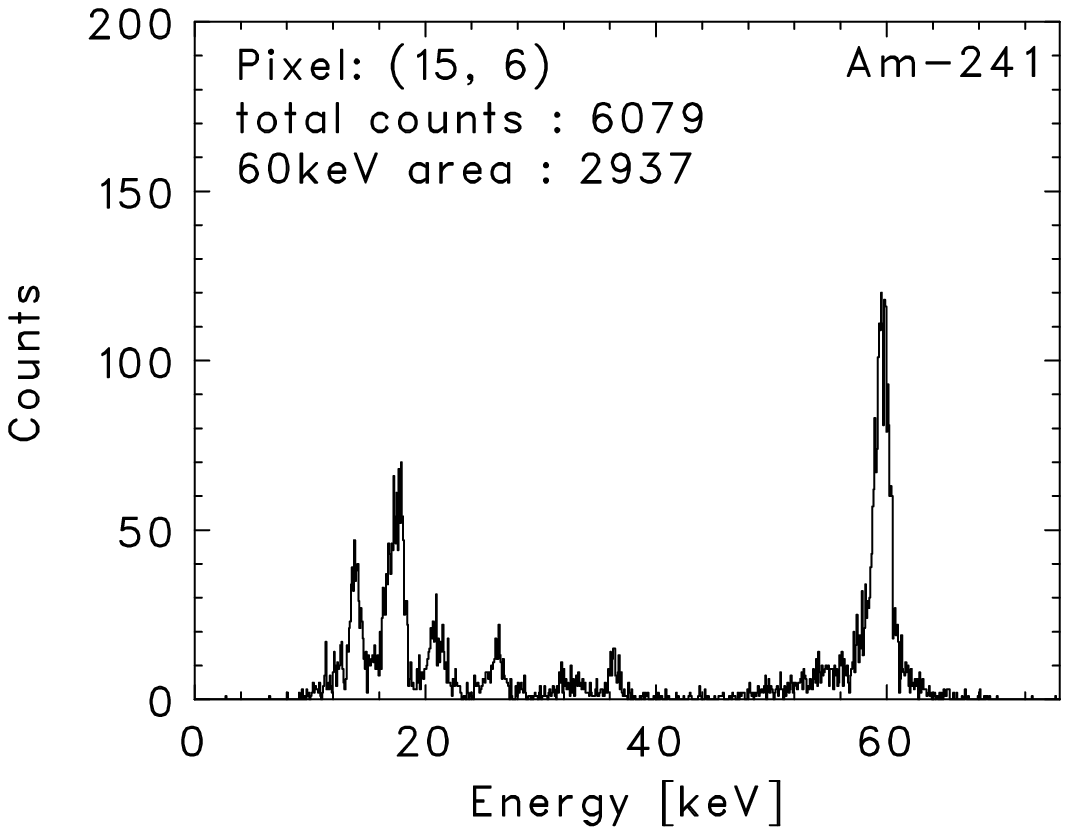}
	\caption{Spectra for the two extremes of the total counts. }
    	\label{fig:spectra for extremes} 
\end{center}
\end{figure}

\begin{figure}[t]
\begin{center}
	\begin{minipage}[t]{14cm}
	\begin{center}
		\includegraphics[keepaspectratio, width = 12cm, clip]{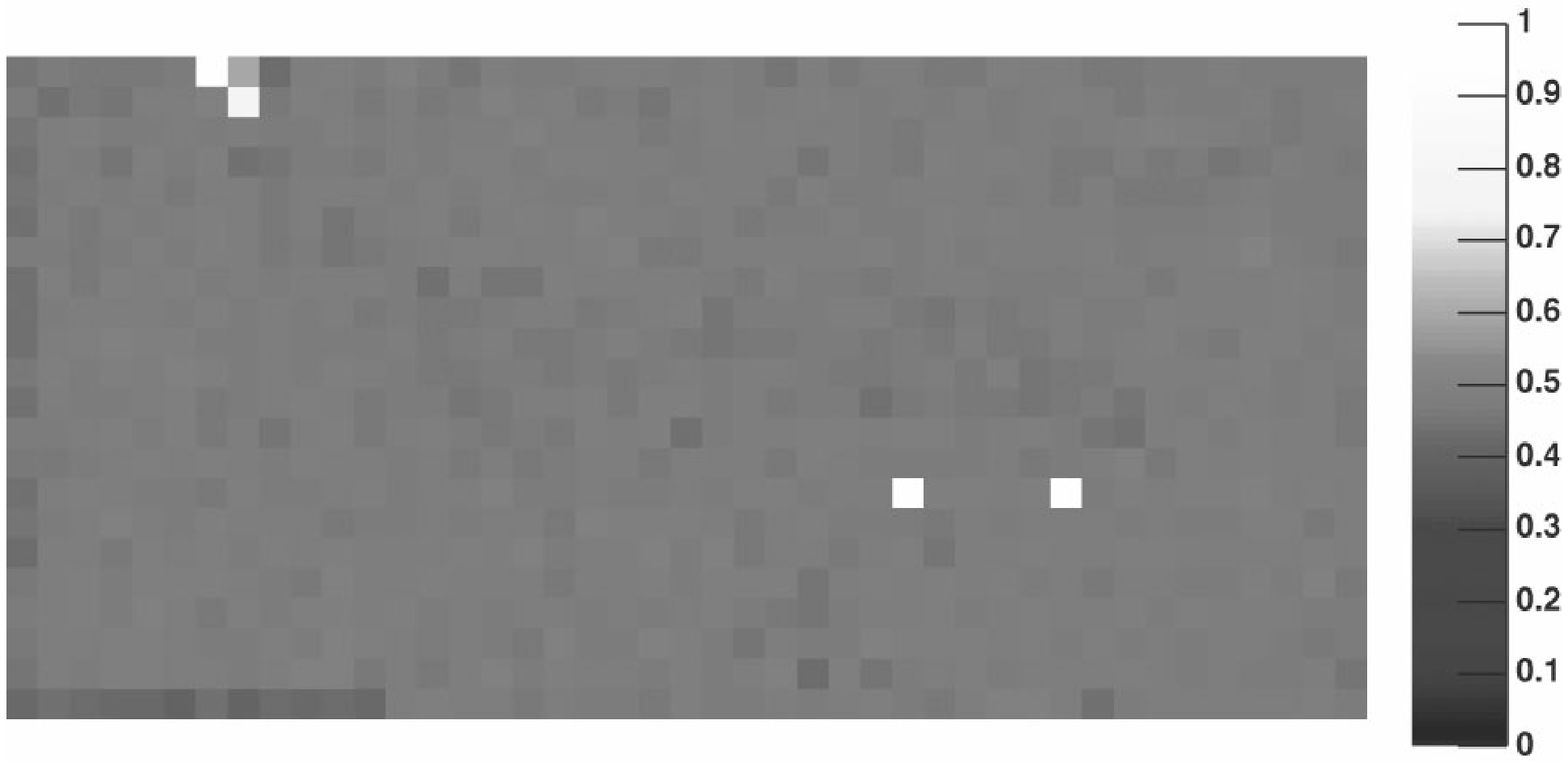}
		\caption{Map for the 60 keV peak area normalized by the total counts detected by each pixel.  The white spots are disabled pixels.}
  		\label{fig:area 2dim} 
   	\end{center}
	\end{minipage}

\vspace{1cm}
	\begin{minipage}[t]{14cm}
	\begin{center}
		\includegraphics[keepaspectratio, width = 9cm, clip]{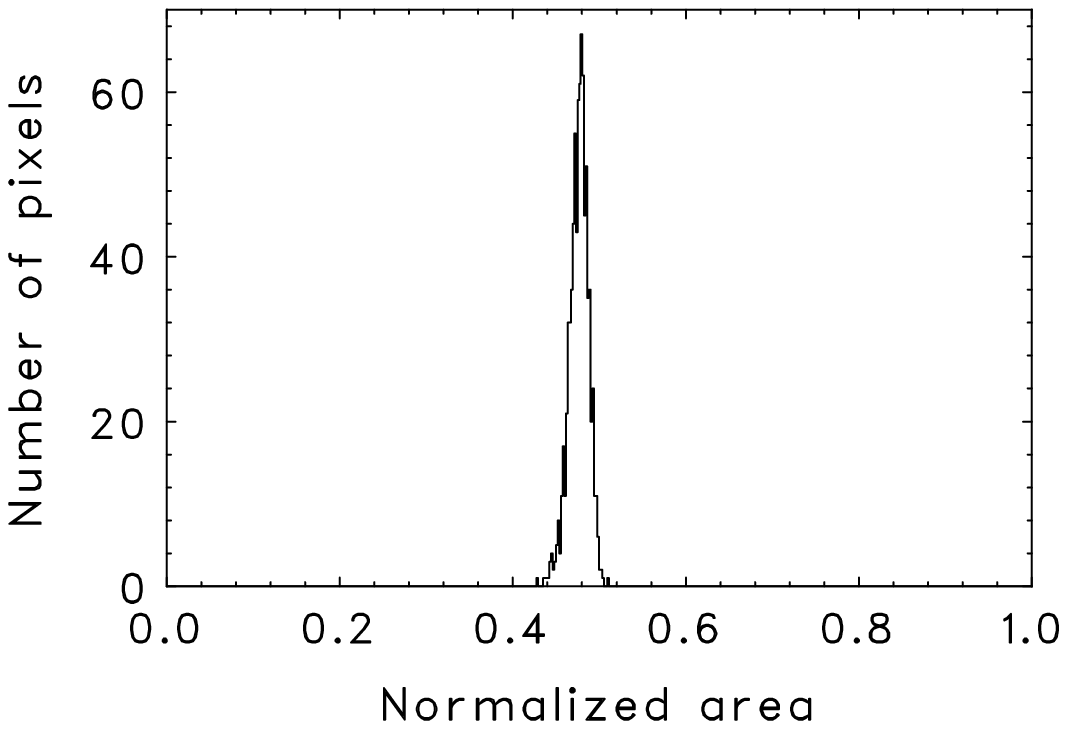}\hspace{1cm}
		\caption{Distribution for the 60 keV peak area normalized by the total counts.  The variation of the distribution is 2.1 \%.} 
	   	\label{fig:area 1dim} 
   	\end{center}
	\end{minipage}
\end{center}
\end{figure}

The variation of the count rate is also an important aspect for the pixel detector, 
since it reflects the variations of the detection efficiency in the detector. 
 In order to produce precise-spectroscopic images, 
 the flat efficiency in the detector plane  is desirable to reduce the systematic errors.
 The uniformity of charge transport properties is again important here. 
 This is because the pixel-to-pixel variation of the fraction of low energy tail affects the flatness
 of the image when we specify the energy band.  
Distribution of the area for the 60 keV peak is plotted in Fig. \ref{fig:60keV counts}.
The area is calculated by integrating counts from 45 keV to 65 keV.
The variation of the distribution is 8.9 \%, which is larger than the statistical variation of 1.1 \%.  
This result is inconsistent with the result for the planar CdTe detector \cite{Ref:Takahashi_IEEE2001,Ref:KN-NIM2003}.

In order to investigate the reason for the variation of the area,
 we compare the counts in the 60 keV peak and the total count, and obtain a scatter plot (Fig. \ref{fig:60keV counts vs total counts}).
From the figure, a strong correlation between the two values is observed.
In Fig. \ref{fig:spectra for extremes}, we compared the actual spectra of the pixels with the highest and the lowest count at 60 keV.  Although the counts are different, the shape of the spectra is very similar.
This similarity implies that the variation of the peak area or the total count rate is not
caused by the spatial inhomogeneity of charge transport properties, such as mobility and lifetime.
This variation then could be explained by two possibilities. 
One is a  reduction of the effective area caused by the bump bonding process.
The other is the dead time variations due to the algorithm currently installed in the
data acquisition (DAQ) system.  We will study this issue more.

If we assume that the variation of total counts is due to the dead-time variations for 
each pixel, we can use it  as an indicator of the actual integration time for each pixel.
Figure \ref{fig:area 2dim} shows a map of the ``normalized" area (area ratio). In the figure,
the integrated counts from 45 keV to 65 keV are normalized by the total counts of the spectrum
for each pixel.  The resultant 
 variation of the area ratio is 2.1 \% while the statistical error is 1.3 \%.

\section{Imaging Spectroscopy} 

\begin{figure}[h]
\begin{center}
	\includegraphics[width= 16cm, keepaspectratio]{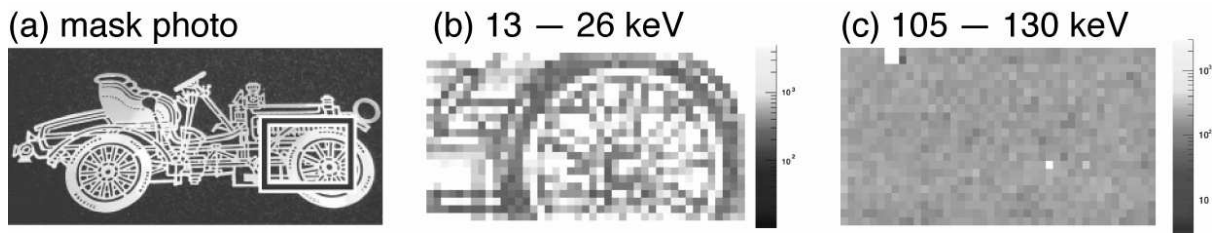}
\end{center}
\caption[] {(a) Photo of a mask, and images using (b) 13 keV - 26 keV lines of $^{241}$Am and (c) the 122 keV line of $^{57}$Co, respectively. The mask is made of brass covered with gold.  A rectangular area is used in the measurement.}
\label{fig:image} 
\end{figure}

We perform imaging measurements to demonstrate the performance of imaging spectroscopy.  An image mask, made of brass covered with gold, is mounted 6 mm above the detector.  It has a length of 70 mm and a thickness of 0.8 mm.  The area marked in Fig. \ref{fig:image}  (a)  is used for the measurement.   $^{241}$Am and $^{57}$Co are located 40 cm above the detector so that the shadow image is obtained with the CdTe pixel detector.  The detector is operated at --20 $^{\circ}$C and 80 V for the pixel is applied.

As shown in Fig. \ref{fig:image} (b) and (c), we have succeeded in imaging spectroscopy using the detector. Fig. \ref{fig:image} (b) is drawn by 13 keV - 26 keV lines from $^{241}$Am, 
while Fig. \ref{fig:image} (c) is by 122 keV line from $^{57}$Co.  
The spoke feature of the mask, which is the same as the pixel size of 500 $\mu$m, is clearly resolved while a featureless image is obtained for the image selected from the 105 -- 130 keV energy band.  
This result is consistent with the fact that the mask is transparent above 100 keV.

\section{SUMMARY}

We have developed a large CdTe pixel detector by utilizing recent technologies of an uniform CdTe single crystal, the two-dimensional ASIC, and the In/Au stud bump bonding.  
When the detector is operated at a temperature of --50 $^{\circ}$C, the energy resolution (FWHM) for the 14 keV and 60 keV $\gamma$-rays is 0.67 keV and 0.89 keV, respectively.
The voltage scanning measurement shows that 10 \% of the total event become multi-pixel events, which contributes to the tail structure in thespectra obtained.  
The detector shows high uniformity as well as the spectral and imaging performances.  
The location of the peak agrees to within 0.82 \% after the analog gain is calibrated by pulser data,
and the average energy resolution is 1.04 keV.
The variation of the 60 keV peak counts is found to be 8.9 \%, which is inconsistent with the result for the planar CdTe detector.  Although the reason for this variation is still under investigation, if we assume a ``normalized'' area, the resultant variation is 2.1 \%.
In the demonstration of imaging spectroscopy, the detector resolves a 0.5 mm wide line, which is the same as the positional resolution.

\acknowledgments     
The authors would like to thank M. Onishi for his dedicated help throughout the detector development.


\end{document}